\documentclass[floats,preprint,superscriptaddress,showpacs,nofootinbib]{revtex4}
\usepackage{bm}
\usepackage{epsfig}
\usepackage{rotating}
\usepackage{amsmath}
\usepackage{amssymb}
\usepackage{cancel}

\newcommand{\nn}{\nonumber}
\newcommand{\beq}{\begin{equation}}
\newcommand{\eeq}{\end{equation}}
\newcommand{\bea}{\begin{eqnarray}}
\newcommand{\eea}{\end{eqnarray}}
\newcommand{\ben}{\begin{eqnarray*}}
\newcommand{\een}{\end{eqnarray*}}

\def\vec#1{{\bf #1}}
\def\D0{D\O}

\newcommand{\Dsl}{\not\!\!D}
\newcommand{\psl}{\not\!p}
\newcommand{\nsl}{\not\!n}

\newcommand{\qsl}{\not\!q}

\newcommand{\bra}[1]{\left\langle #1 \right|}
\newcommand{\ket}[1]{\left|#1\right\rangle}

\newcommand{\V}[1]{\mathbf{#1}}

\newcommand{\dfs}{d}
\newcommand{\s}[1]{\rlap \slash #1}
\newcommand{\ps}{\s{p}}

\newcommand{\ns}{\s{n}}
\newcommand{\pars}{\s{\partial}}
\newcommand{\nbs}{\s{\bar{n}}}

\renewcommand{\d}{\partial}

\begin{document}

\title{Implications of $SU(2)_L \times U(1)$ Symmetry for SIM(2) Invariant Neutrino Masses}%
 
\author{Alan Dunn}%
\email{amd34@phy.duke.edu}
\affiliation{Department of Physics, Duke University, Durham NC 27708, USA}
 
\author{Thomas Mehen}%
\email{mehen@phy.duke.edu}
\affiliation{Department of Physics, Duke University, Durham NC 27708, USA}
\affiliation{Jefferson Laboratory, 12000 Jefferson Ave., Newport News VA 23606}

\date{\today}

\begin{abstract}
 
We consider $SU(2)_L \times U(1)$ gauge invariant generalizations of a nonlocal, Lorentz violating mass term for
neutrinos that preserves a SIM(2) subgroup. This induces Lorentz violating  effects in QED as well as  tree-level 
lepton family number violating interactions. Measurements of $g_e-2$ with trapped electrons  severely constrain
possible SIM(2) mass terms for electrons which violate C invariance.   We study Lorentz violating effects in a
C invariant and SIM(2) invariant extension of QED.  We examine the  Lorentz violating interactions of nonrelativistic
electrons with electromagnetic fields to determine their impact on the spectroscopy of hydrogen-like  atoms and
$g_e-2$ measurements with trapped electrons. Generically, Lorentz violating corrections are suppressed by
$m_\nu^2/m_e^2$ and are within  experimental limits.  We study one-loop corrections to electron and photon
self-energies and point out the need  for a prescription to handle IR divergences induced by the nonlocality of the
theory. We also calculate the tree level contribution to $\mu \to e + \gamma$ from SIM(2) invariant mass terms.

\end{abstract}

\maketitle

In very special relativity (VSR)~\cite{Cohen:2006ky}, the laws of nature are not invariant under the  full Lorentz
group but only a subgroup called SIM(2). In SIM(2) invariant theories there is  a preferred  light-like
four-vector, called $n^\mu$. The unbroken generators of SIM(2) consist of the generators of little group of
$n^\mu$ as well as boosts under which $n^\mu$ transforms as $n^\mu \to e^\alpha n^\mu$. SIM(2) invariance is
sufficient to guarantee that particles obey the relativistic  dispersion relation, $E^2 = p^2 + m^2$. This ensures
that SIM(2) invariant  theories will satisfy constraints that come from relativistic kinematics, so the violation of Lorentz 
invariance in these theories is more subtle. Since SIM(2) has only one-dimensional representations, states which
are degenerate in a Lorentz invariant theory can split into nondegenerate states which satisfy relativistic
dispersion relations with different masses. For instance, we will see below  that for certain kinds of SIM(2)
invariant mass terms, the four spin states of the electron and positron split into two nondegenerate pairs of
states. SIM(2) theories respect CPT~\cite{Cohen:2006ky}, which accounts for the remaining degeneracy. The other
way the SIM(2) invariant theories can violate Lorentz invariance is via interactions. For example, the couplings
of an electron to background electromagnetic fields, particle decay rates, and cross sections can depend on the
preferred four-vector $n^\mu$ and therefore violate Lorentz symmetry. These effects cannot be accessed simply by
examining free particle propagation. 

When the SIM(2) generators are combined with parity operation, the full Lorentz group is obtained. Therefore, the breaking of
Lorentz invariance is tied to the breaking of parity. This may naturally suppress Lorentz violating effects. Another
important feature of the SIM(2) group is that it has no invariant tensors, implying that the usual spurion analysis of
Lorentz violation does not apply to SIM(2) invariant theories~\cite{Cohen:2006ky}. In traditional analyses of Lorentz
violation, see e.g., Refs.~\cite{Coleman:1998ti,Coleman:1997xq,Colladay:1996iz,Colladay:1998fq}, Lorentz noninvariant tensors
are introduced and then {\it local} dimension-3 and dimension-4 operators in the  Standard Model containing these tensors are
added to the Standard Model Lagrangian. This approach does not lead to SIM(2) invariant theories, since the  terms with
Lorentz noninvariant tensors will also violate SIM(2) invariance. In SIM(2) invariant theories, Lorentz violating terms in
the Lagrangian are necessarily {\it nonlocal}. 

Finally, SIM(2) invariant theories provide a novel way of generating a mass for a two-component spinor~\cite{Cohen:2006ir}.  This is
especially interesting for theories of the neutrino mass. To give neutrinos a mass in the Standard Model one must either add
a gauge singlet right-handed neutrino to form a Dirac mass term or else add a Majorana mass term which violates lepton
number. SIM(2) invariant theories allow for a novel mass term which requires neither a right-handed neutrino nor lepton
number violating interactions. Ref.~\cite{Cohen:2006ir} proposed a SIM(2)  invariant mass term for neutrinos and studied its
effect on the electron spectrum in tritium beta-decay.

When the SIM(2) invariant mass term of Ref.~\cite{Cohen:2006ir} is made  invariant under the $SU(2)_L \times
U(1)$ gauge group of the Standard Model, Lorentz violating interactions will be induced in other sectors of the
theory, including QED. The purpose of this paper is
to study the Lorentz violating effects implied by the SIM(2) invariant neutrino mass.  We also make a preliminary
investigation of loop effects in a SIM(2) invariant extension of QED. 

The paper is organized as follows. In the section I, we generalize the mass term for neutrinos considered in
Ref.~\cite{Cohen:2006ir} by requiring the theory to be  invariant under the $SU(2)_L \times U(1)$ gauge symmetry of
the Standard Model. This leads to Lorentz violating effects in QED as well as tree-level lepton family number violating
interactions.  In section II, we focus on a C invariant and SIM(2) invariant version of QED, called VSRQED, and derive
a nonrelativistic Hamiltonian for electrons in the presence of background electric and magnetic fields. This
Hamiltonian is used to determine the consequences of VSRQED for tests of Lorentz invariance with trapped electrons.
VSRQED corrections to the spectroscopy of hydrogenic atoms are also considered. Since Lorentz violating
effects in this theory are suppressed by $m_\nu^2/m_e^2$, the predicted Lorentz violating effects are tiny and within current
experimental bounds. In section III, we study one-loop self-energy corrections to the photon and electron
propagators. The nonlocality of the SIM(2) invariant theory leads to IR divergences  which
make the loop integrals ill-defined. We propose a prescription for dealing with these IR divergences. Evaluated
with this prescription, the loop diagrams  give corrections that satisfy the Ward Identity. In addition,  the
photon remains massless at one loop  and no Lorentz violating modifications to the photon propagator are induced.
Furthermore, the one-loop correction to the SIM(2) invariant electron mass term is finite. In section IV, we calculate
the tree level contribution to the lepton family number violating decay  $\mu \to e + \gamma$. Our results are
summarized in section V. In Appendix A, we derive Feynman rules for VSRQED.

\section{SIM(2) Invariant Modifications of the Standard Model}

The SIM(2) invariant, Lorentz violating, modified Dirac equation
for the left-handed neutrino proposed in 
Ref.~\cite{Cohen:2006ir} is 
\begin{equation}
\left(\ps - \frac{m_\nu^2}{2} \frac{\ns}{n \cdot p} \right) \nu_L(p) = 0 \, .
\label{eq:eqmn}
\end{equation}
Though Eq.~(\ref{eq:eqmn})   violates Lorentz invariance, 
by squaring the modified Dirac operator it is obvious that 
\bea
(p^2-m_\nu^2 ) \nu_L(p) = 0 \, ,
\eea
so the left-handed neutrino has a relativistic, massive dispersion relation. Thus, SIM(2) invariance provides
an alternative way to introduce a neutrino mass without adding either a Dirac or Majorana mass term.

The Lagrangian that would give rise to Eq.~(\ref{eq:eqmn}) is 
\bea\label{nulag}
{\cal L} = \bar \nu_L \left( i \pars -\frac{m_\nu^2}{2}\frac{\ns}{i n\cdot \partial} \right) \nu_L \, , 
\eea
where $\nu_L$ is the left-handed neutrino field. The second term in the Lagrangian is nonlocal and has a structure similar 
to kinetic terms in the light-cone QCD Hamiltonian~\cite{Wilson:1994fk,Brodsky:1997de} as well as Soft-Collinear Effective
Theory  (SCET)~\cite{Bauer:2000yr,Bauer:2001yt}. The action is ill-defined  for modes  of $\nu_L$ satisfying $i n \cdot
\partial \nu_L =0$.  In the tree level calculations of this paper we will not have to worry about this issue, however, in
loop interactions this leads to infrared divergences and a prescription is needed to define the theory.  We postpone
discussion of this issue until we study loop diagrams in section \ref{sec:loops}. 

In this section we focus on the implications of embedding the Lagrangian of  Eq.~(\ref{nulag})  in the Standard Model. A necessary
requirement is that the action be invariant under the $SU(2)_L \times U(1)$ gauge group. Eq.~(\ref{nulag}) can be made gauge invariant by
replacing the neutrino in Eq.~(\ref{nulag}) with the left-handed lepton doublet and making the derivatives in Eq.~(\ref{nulag}) covariant. We
can also consider a similar term for the right-handed electron field. We will restrict our attention to the lepton sector of the Standard
Model. For three generations of leptons, we obtain  the following SIM(2) invariant mass terms,
\begin{equation}
\mathcal{L}_{SIM(2)} = - \frac{1}{2} \bar{l}_{L,i} \, M^2_{i j} \, \frac{\ns}{i n \cdot D} l_{L,j} - \frac{1}{2} \bar{e}_{R,i} \,
\tilde M^2_{i j} \, \frac{\ns}{i n \cdot D} e_{R,j} \, ,
\end{equation}
where the  $SU(2)_L$ doublets are denoted by  $l_{L}$, $i$ and $j$ are family indices, and $M^2$ and $\tilde M^2$ 
are Hermitian matrices. Summation over repeated indices is implied. Once the electroweak symmetry is broken the mass terms in 
the theory are
\bea
{\cal L} &=& -v (\lambda_{i j}\, \bar e_{L,i} \, e_{R,j} + h.c.) - \frac{1}{2}\bar \nu_{L,i} \, M^2_{ij} \, \frac{\ns}{i n\cdot D} \nu_{L,j}
\nn \\ &&- \frac{1}{2}\bar e_{L,i} \, M^2_{ij} \, \frac{\ns}{i n\cdot D} e_{L,j}
- \frac{1}{2}\bar e_{R,i}\,  \tilde M^2_{ij} \, \frac{\ns}{i n\cdot D} e_{R,j} \, ,
\eea
where $e_{L,i}$ and $e_{R,i}$ are the left-handed and right-handed charged lepton fields, respectively, 
$v$ is the Higgs expectation value, and the $\lambda_{ij}$ are the usual Standard Model Yukawa couplings. 
Three independent unitary transformations in family space, 
\begin{equation}
e_{L} \rightarrow U_L \, e_L, \qquad e_{R} \rightarrow U_R \,e_R, \qquad\nu_L \rightarrow V \,\nu_L \, ,
\end{equation} 
can be used to diagonalize some of the mass terms.
The matrix $V_M  = U_L^\dagger V$ is the neutrino mixing matrix.  Not all SIM(2) invariant mass terms can be simultaneously diagonalized. 
Choosing $U_L$ and $U_R$ to diagonalize the charged lepton Dirac  mass matrix and $V$ to diagonalize the neutrino mass matrix, we find
\bea\label{genL}
{\cal L} &=& -  m_{\ell,i} \, (\bar e_{L,i}\,  e_{R,i} + h.c. )
-  \frac{1}{2}\bar \nu_{L,i}\, m^2_{\nu,ii} \,\frac{\ns}{i n\cdot D} \nu_{L,i}
\nn \\ && - \frac{1}{2}\bar e_{L,i} \,(V_M \, m^2_{\nu} \,V^\dagger_M)_{ij} \,\frac{\ns}{i n\cdot D} \,e_{L,j}
- \frac{1}{2}\bar e_{R,i} \,(U_R^\dagger \, \tilde M^{2} \, U_R)_{ij}\, \frac{\ns}{i n\cdot D} \,e_{R,j} \, ,
\eea
where $m_{\ell,i}$ would be the charged lepton masses in the absence of SIM(2) violating terms, and we have defined the
diagonal matrix  of neutrino masses squared to be $m_\nu^2 = V^\dagger M^2 V$. The SIM(2) invariant mass matrix for the 
left-handed leptons is determined in terms of  the neutrino masses and mixing matrix. It is nondiagonal and therefore these
terms introduce additional lepton family number violation. The SIM(2) invariant mass matrix for the right-handed leptons is
undetermined and should generically be nondiagonal since the matrix $U_R$ which is needed to diagonalize the Yukawa
couplings, $\lambda_{ij}$, is in general not the same as the one required to diagonalize $\tilde M^{2}_{ij}$. 

The existence of different SIM(2) invariant mass matrices for the left-handed and right-handed leptons leads to 
a wave equation for charged leptons that is  more complicated than the neutrino's wave equation. In particular it 
can lead to a breaking of the degeneracy of the spin states of the charged leptons. We will focus on the
electron since we are primarily interested in looking for Lorentz violating effects in SIM(2) invariant
extensions of QED. Lorentz violating effects which split the degeneracy of the electron's spin states 
along a preferred direction in space are strongly constrained by modern experiments with trapped electrons~\cite{Mittleman:1999it}. 
 
For the electron the 
modified Dirac equation takes the form 
\bea\label{ewe}
\left(\ps - m_0 - \frac{1}{2} \frac{\ns}{n \cdot p} (M^2_+ + M^2_- \gamma_5)\right)u(p) = 0 \, , 
\eea 
where $u(p)$ is electron spinor, $m_0 = m_{\ell,1}$, and
$M^2_\pm =\pm \frac{1}{2}(V_M m^2_\nu V_M^\dagger)_{11} + \frac{1}{2}(U_R^\dagger \tilde M^2 U_R)_{11}$. 
Note that the combination $M^2_+ - M^2_- = (V_M m^2_\nu V_M^\dagger)_{11}$ is the so-called electron based neutrino
mass squared, $m_\beta^2$. The orthogonal linear combination  $M^2_+ + M^2_- = (U_R^\dagger \tilde M^2 U_R)_{11}$  is not determined by parameters 
in the neutrino sector.

The current constraint on $m_\beta$ from tritium beta decay experiments is 
$m_\beta \, <\, 2 \,{\rm eV}$~\cite{PDG:2006}. This constraint assumes a Lorentz invariant neutrino mass.
Ref.~\cite{Cohen:2006ir} pointed out that the effect of a SIM(2) invariant neutrino mass on the beta decay 
spectrum is roughly equivalent to that of a Lorentz invariant neutrino mass that is a factor of two more massive.
So for SIM(2)
invariant masses, the current constraint from tritium beta decay should be interpreted as $m_\beta < 1 \, {\rm eV}$. 
Cosmology gives a comparable constraint on the sum of neutrino masses, 
which must be greater than $m_\beta$. The bounds are somewhat model dependent, with quoted lower bounds ranging from 
$0.6 - 2.0 \, {\rm eV}$~\cite{Crotty:2004,Barger:2004,Ichikawa:2005}. We do not know how the cosmology
bounds would be affected if the neutrino masses are SIM(2) invariant rather than Lorentz invariant.
We will assume $m_\beta < 1$ eV for the remainder of the paper.

To determine the mass spectrum, we invert the operator in Eq.~(\ref{ewe}) and find the locations
of the poles in the electron propagator. First we define
\begin{equation}
p'^\mu \equiv p^\mu - \frac{M^2_+}{2} \frac{n^\mu}{n \cdot p} \, ,
\end{equation}
and note that $n \cdot p = n \cdot p^\prime$ and $p'^{\,2} = p^2 -M^2_+$.
Left multiplication of the operator in Eq.~(\ref{ewe}) by $\ps^{\, \prime} - m_0 + \frac{M_-^2}{2} \frac{\ns}{n \cdot p} \gamma_5$
yields
\begin{equation}
p'^{\, 2} + m_0^2 - 2 m_0 \ps^{\, \prime} - M_-^2 \gamma_5 \, .
\end{equation}
Next, multiplying by $2 m_0 \ps^{\, \prime} + (p'^{\, 2} + m_0^2) + M_-^2 \gamma_5$ yields the operator
\bea
(p'^{\, 2} - m_0^2)^2 -  M_-^4\, .
\eea
This sequence of operations shows that the electron propagator is 
\begin{equation}\label{inv}
i\, \frac{(2 m_0 \ps -  m_0 M_+^2 \frac{\ns}{n \cdot p} + (p^2 + m_0^2 - M_+^2) + M_-^2 \gamma_5)
(\ps - m_0 - \frac{1}{2} \frac{\ns}{n \cdot p}(M^2_+ - M_-^2 \gamma_5))}{(p^2 - m_0^2 - M_+^2)^2 -  M_-^4} \, ,
\end{equation} 
so the electron propagator will have poles at
\begin{equation}
p^2 = E_{\pm}^2 \equiv m_0^2 + M_+^2 \pm  M_-^2 \, .
\end{equation}
The four spin states of the electron and positron are  degenerate only when $M_- = 0$. 

To better understand how the states are split we study the solutions to the modified 
Dirac equation. We work in the rest frame $p^\mu =(E,\V{0})$, and choose
the $z$-axis to be the spatial direction singled out by the four-vector, $n^\mu=(1,0,0,1)$. Then the modified 
Dirac equation is
\bea
\left(E \, \gamma^0 -m_0 -\frac{1}{2E}(\gamma^0-\gamma^3)(M_+^2+M_-^2\gamma_5)\right) u(p)=0\ \, .
\eea
Using the Bjorken-Drell conventions for the gamma matrices, we find the following solutions:
\bea
 u_\downarrow\left(E_+\right) = \left(\begin{array}{c} 0 \\ 1 \\  0 \\ \frac{M_-^2+M_+^2}{(m_0+E_+)^2} \end{array}\right)
\qquad 
v_\downarrow\left(-E_+\right) = \left(\begin{array}{c} 0 \\  \frac{M_-^2+M_+^2}{(m_0+E_+)^2} \\  0 \\ 1\end{array}\right) 
\nn \\\nn \\
u_\uparrow\left(E_-\right) = \left(\begin{array}{c} 1 \\ 0 \\  \frac{M_-^2 - M_+^2}{(m_0+E_-)^2} \\ 0 \end{array}\right)
\qquad 
v_\uparrow\left(-E_-\right)= \left(\begin{array}{c} \frac{M_-^2 - M_+^2}{(m_0+E_-)^2} \\  0 \\ 1  \\ 0\end{array}\right)
\nn \, , 
\eea
where $E_\pm = \sqrt{m_0^2 + M_+^2 \pm M_-^2}$. We have labeled the positive energy spinors $u_\uparrow$ and
$u_\downarrow$, and the negative energy spinors  $v_\uparrow$ and $v_\downarrow$. In the absence of SIM(2) invariant  mass
terms, $u_\downarrow \,(u_\uparrow)$  corresponds to an electron with spin-down (spin-up) along the $z$-axis.  The
negative energy solutions are interpreted as antiparticle states which  are degenerate with the  particle states, as
expected because the SIM(2) invariant masses  respect CPT. We see that when $M_- \neq 0$ electrons with spin-up and
spin-down along the $z$-axis are split by $E_+ - E_- \approx M_-^2/(2 m_0)$. 

Such a splitting will lead to a diurnal variation of the anomalous precession frequency of trapped  electrons in experiments that 
measure $g_e-2$.  Ref.~\cite{Mittleman:1999it} places bounds on diurnal variation of the anomalous precession.
Ref.~\cite{Mittleman:1999it} analyzes the data using the formalism of Ref.~\cite{Bluhm:1997ci}, which considers an electron wave equation of the form:
\bea\label{lvde}
(\ps - m_e - e A_\mu \gamma^\mu -a_\mu \gamma^\mu -  b_\mu \gamma_5\gamma^\mu) u(p) = 0 \, ,
\eea
where $A_\mu$ is the electromagnetic vector potential, and $a_\mu$ and $b_\mu$ are background four-vectors.  The vector $a_\mu$ has no physical effects since 
it can be removed by the gauge transformation $A_\mu \to A_\mu -\partial_\mu \Lambda$, where $\Lambda= a_\mu x^\mu/e$. The four-vector $b_\mu$ gives rise to a
diurnal variation of the  anomalous precession frequency ~\cite{Bluhm:1997ci,Mittleman:1999it}. Eq.~(\ref{lvde}) does not have the same form as Eq.~(\ref{ewe}).
Invariance under $n^\mu \to e^\alpha \, n^\mu$ allows us to rescale $n^\mu$ so that $n^\mu =(1, \V{n})$, where $\V{n}$ is a spatial unit vector. Then for
nonrelativistic electrons  we can approximate  $n\cdot p \approx m_e$ and  identify $a_\mu = M_+^2/(2 m_e)\, n_\mu$ and $b_\mu = M_-^2/(2 m_e) \, n_\mu$. The
observation of Ref.~\cite{Bluhm:1997ci} that $a_\mu$ does not affect the precession frequency is consistent with our observation that the term proportional to
$M_+^2$ does not split the electron spin states, and the diurnal variation of the anomalous precession due to $b_\mu$ found in Ref.~\cite{Bluhm:1997ci} is
consistent with the splitting of electron spin states we find in the presence of the term  proportional to $M_-^2$. Ref.~\cite{Mittleman:1999it} obtains a bound
on a product of a certain component of $b^\mu$ times geometrical factors associated with the orientation of the experiment relative to the background vector.
Assuming these geometrical factors are of order unity, we find $M_-^2 = -\frac{1}{2}(V_M m^2_\nu V_M^\dagger)_{11} + \frac{1}{2}(U_R^\dagger \tilde M^2
U_R)_{11} \lesssim 10^{-10}\, {\rm eV}^2$. This value of $M_-$ appears to require considerable fine tuning since $(V_M m_\nu^2 V_M^\dagger)_{11} =
m_\beta^2 \gtrsim 10^{-5}$ eV$^2$, as we will later show.

However, it is possible to enforce $M_- = 0$ by imposing charge conjugation invariance
on the SIM(2) invariant theory. The action of discrete symmetries 
on the terms in the Lagrangian which give rise to the SIM(2) invariant mass terms of Eq.~(\ref{ewe})  
is given in Table~\ref{discrete}. Here $\bar{n}^\mu$ is the parity image of $n^\mu$, i.e.,
if $n^\mu =(1,\vec{n})$ then $\bar{n}^\mu=(1,-\vec{n})$. 
\begin{table}[t!] 
  \begin{tabular}{ccc|ccc|ccc|ccc}
& & & & C &  & & P & & & T \\
\hline\hline
& & & & & & & & & & & \\
&$ \bar{\psi} \ns  \frac{1}{i n \cdot \d} \psi(t, \V{x})$ & &
&$\bar{\psi} \ns \frac{1}{i n \cdot \d} \psi(t, \V{x})$ & &
&$\bar{\psi} \nbs  \frac{1}{i \bar{n} \cdot \d} \psi(t, -\V{x})$ & &
&$\bar{\psi} \nbs  \frac{1}{i \bar{n} \cdot \d} \psi(- t, \V{x})$ & \\
& & & & & & & & & & & \\
\hline
& & & & & & & & & & & \\
&$ \bar{\psi} \ns \gamma_5 \frac{1}{i n \cdot \d} \psi(t, \V{x})$ & &
&$- \bar{\psi} \ns \gamma_5\frac{1}{i n \cdot \d} \psi(t, \V{x})$ & &
&$- \bar{\psi} \nbs \gamma_5 \frac{1}{i \bar{n} \cdot \d} \psi(t, -\V{x})$ & &
&$\bar{\psi} \nbs \gamma_5 \frac{1}{i \bar{n} \cdot \d} \psi(- t, \V{x})$ & \\
& & & & & & & & & & & \\
\end{tabular}
\caption{Action of discrete symmetries on SIM(2) invariant mass terms.}
\label{discrete}
\end{table}
From Table~\ref{discrete} we see that both SIM(2) invariant mass terms respect
CPT, which is consistent with the arguments of Ref.~\cite{Cohen:2006ir}. The mass term without $\gamma_5$
is also invariant under C and PT, while the mass term with $\gamma_5$ is not invariant under either.
Thus, if we impose C invariance, or equivalently PT invariance, we can get rid of terms proportional
to $M_-^2$. For the case $M_-^2 = 0$, the electron propagator is 
\begin{equation}\label{cinv}
i \, \frac{\ps + m_0 - \frac{M^2_+}{2} \frac{\ns}{n \cdot p}}{p^2 - m_0^2 - M_+^2}  \, ,
\end{equation}
which is considerably simpler than the electron propagator in Eq.~(\ref{inv}).
Thus, imposing $C$ invariance
makes the SIM(2) invariant version of QED easier to analyze. 
 
There are two important  conclusions we wish to draw from the analysis of this section. One is that SIM(2) invariant
neutrino masses necessarily require SIM(2) invariant mass terms for the electron, implying  Lorentz violating effects in QED. Tests of
Lorentz invariance in QED have reached a high level of precision so the hypothesis that the neutrino has a SIM(2) invariant mass term
can be tested by looking for Lorentz violating effects in this sector of the Standard Model. Current experiments with trapped electrons force one
to consider a C invariant and  SIM(2) invariant extension of QED, which we will call VSRQED.
This theory is studied in the next two sections of the paper. The second important point is that the
SIM(2) invariant  mass terms in the lepton sector are family  non-diagonal so  SIM(2) invariant mass terms for neutrinos imply new tree-level
contributions to lepton family number violating   processes   such as  $\mu \to e + \gamma$. This process is studied in section IV.

\section{Non-Relativistic Electrons in VSRQED}

In this section we derive the coupling of nonrelativistic electrons to electric and magnetic fields and study the effect these 
interactions have on measurements of $g_e-2$ using trapped atoms and on the 
spectroscopy of hydrogenic atoms. We will impose C invariance so the  modified Dirac equation for the electron is
\bea \label{vsrmde}
\left( \psl - m_0 - \frac{m_\beta^2}{2} \frac{\nsl}{n \cdot p} \right) u_n(p) = 0  \, ,
\eea 
where $u_n(p)$ is the electron spinor and $m_\beta^2 = (V_M m_\nu^2 V_M^\dagger)_{11}$
is the electron based neutrino mass. For the remainder of this paper, $u_n(p)$ will denote a
positive energy solution to the modified Dirac equation in Eq.~(\ref{vsrmde}), and $u(p)$ 
will denote will denote a positive energy solution to the unmodified Dirac equation.
It is useful to write the solution $u_n(p)$ in terms of  $u(p)$, 
\bea\label{sln}
u_n(p) &=& \left( 1 - \frac{m_e -m_0}{2}\frac{\nsl}{n\cdot p}\right) u(p) \nn\\
&\approx& \left( 1 - \frac{m_\beta^2}{4 m_e}\frac{\nsl}{n\cdot p}\right) u(p) \, .
\eea 
Note that $m_e = \sqrt{m_0^2 + m_\beta^2}$, $(\psl - m_e )u(p) = 0$, and in the last line we have expanded 
to first order in $m_\beta^2/m_e^2$. This allows us to write the interaction between the electron and 
the electromagnetic fields to $O(m_\beta^2/m_e^2)$ in terms of solutions to the conventional Dirac equation.
If we work in $n\cdot A =0 $ gauge, the Feynman rule for the interaction between the electron and the photon 
is 
\begin{align*}
-i e \bar{u}_n(p') \gamma^\mu u_n(p) \tilde{A}_\mu(q) = -i e \bar{u}(p') \gamma^\mu u(p) \tilde{A}_\mu(q)+ 
i\frac{e}{4} \frac{m_\beta^2}{m_e}\, \bar{u}(p')
 \frac{n_\nu n^\alpha \sigma_{\alpha \mu} \tilde{F}^{\mu \nu}(q)}{(n \cdot p') (n \cdot p)} u(p)+ ... \, .
\end{align*}
For nonrelativistic electrons, we can expand to lowest order in the electron momentum and express the interaction in terms
of two-component  non-relativistic spinors. To lowest order in the electron momentum, we find
\begin{align*}
i\frac{e}{4} \frac{m_\beta^2}{m_e} \, \bar{u}(p') \frac{n_\nu n^\alpha \sigma_{\alpha \mu} \tilde{F}^{\mu \nu}(q)}{(n \cdot p') (n \cdot p)} u(p)
\approx i\frac{e}{2} \frac{m_\beta^2}{m_e^3} \, \psi^\dagger ((\V{n} \times \V{E}) \cdot \V{\sigma} - \V{\sigma} \cdot \V{B} 
+ (\V{n} \cdot \V{\sigma}) (\V{n} \cdot \V{B}))\psi \, .
\end{align*}
We have set  $n^\mu =( 1, \V{n})$, where $\V{n}$ is a unit vector. 
This leads to the following contributions to the Hamiltonian for a nonrelativistic electron in the presence of $\V{E}$ and
$\V{B}$ fields:
\bea\label{nrham}
H _{SIM(2)} = \epsilon \, \mu_B \left[- (\V{n} \times \V{E}) \cdot \V{\sigma} + \V{\sigma} \cdot \V{B} 
- (\V{n} \cdot \V{\sigma}) (\V{n} \cdot \V{B})\right] \, ,
\eea
where $\mu_B = e \hbar/(2 m_e c)$ and $\epsilon = m_\beta^2/m_e^2$. Currently, $\epsilon$ is constrained to lie within the range:
\bea\label{epsrange}
9.2 \, 10^{-17} < \epsilon < 3.8 \, 10^{-12} \, .
\eea
The upper bound comes from  $m_\beta \leq 1\, {\rm eV}$~\cite{PDG:2006}, and 
the lower bound comes from noting that 
\bea
m_\beta^2 =(V_M m_\nu^2 V_M^\dagger)_{11} &=& c_{12}^2 c_{13}^2 m_1^2 + s_{12}^2 c_{13}^2 m_2^2 +  s_{13}^2 m_3^2  \nn \\
&=&   m_1^2 + s_{12}^2 c_{13}^2 \Delta m_{21}^2 +  s_{13}^2 \Delta m_{31}^2 \, .
\eea
Here $c_{ij}= \cos(\theta_{ij}), s_{ij}= \sin(\theta_{ij})$ and $\Delta m_{ij}^2 = m_i^2-m_j^2$. This is a sum of positive definite terms. With $\Delta m_{21}^2 = 8.0 \, 10^{-5} \, {\rm eV}^2$, $s_{12}^2 = 0.31$ and $c_{13}^2 \geq 0.95$,
we find $m_\beta^2 \geq  s_{12}^2 c_{13}^2 \Delta m_{21}^2 \approx 2.4 \, 10^{-5} \,{\rm eV}^2$. There are no lower bounds
on $m_1^2$ or $s_{13}$. 

For nonrelativistic electrons the  SIM(2) invariant mass term gives rise to a Lorentz invariant contribution to $g_e-2$ as well
as Lorentz violating couplings of the electron to background electric and magnetic fields. These interactions  
depend on the preferred direction, $\V{n}$. Note that manifest invariance under the SIM(2) transformation
$n^\mu \to e^\alpha n^\mu$ is lost when the nonrelativistic limit is taken. The unit vector, $\V{n}$, points
 in the preferred direction as seen from the electron rest frame. 
The size of the Lorentz violating corrections are fixed by parameters in the neutrino sector. Therefore,
the hypothesis that neutrino  mass terms are SIM(2) invariant rather then Lorentz invariant   predicts tiny Lorentz violating 
effects in QED whose size is determined by $\epsilon$. Tests of Lorentz and rotational invariance in QED 
which constrain $\epsilon$ to be less than $9 \, 10^{-17}$ will rule out the SIM(2) invariant neutrino mass hypothesis.
Observation of Lorentz violation in the range predicted by Eq.~(\ref{nrham}) and Eq.(\ref{epsrange}) would provide support for 
SIM(2) invariant neutrino masses.
We will consider the effect of the terms in Eq.~(\ref{nrham}) on experiments that search for 
an electric dipole moment (edm) of the electron,  experiments
with trapped electrons,  and the spectroscopy of hydrogenic atoms.

The first Lorentz violating term in Eq.~(\ref{nrham}) is like an electron edm, except the electron spin
couples to $\V{n} \times \V{E}$ instead of $\V{E}$. Searches for an electron edm use an apparatus 
in which atomic beams are passed through regions in which $\V{E}$  and $\V{B}$ fields are either aligned or antialigned,
and then an edm will result in a splitting between the atomic levels in the two beams, which can be measured
via interference~\cite{Regan:2002}. These experiments are insensitive to the $\V{n} \times \V{E}\cdot\V{\sigma}$ interaction in 
Eq.~(\ref{nrham}) because the interaction vanishes when $\V{E}$ and $\V{B}$ are aligned. We are not aware of experiments 
which look for splittings between spin states of atoms in {\it crossed} $\V{E}$ and $\V{B}$ fields, which is what would 
be required to test for interactions proportional to $\V{n} \times \V{E}\cdot\V{\sigma}$. Note that the current bound 
on the electron edm is $|d_e| \leq 1.6 \, 10^{-27}$ $e$-cm $= 8.3 \, 10^{-17} \, \mu_B$. Thus a bound on 
$\epsilon \, \mu_B$ comparable in size to present bounds on the electron edm requires $\epsilon$ smaller
than the lower bound in Eq.~(\ref{epsrange}) and would rule out the SIM(2) invariant neutrino mass hypothesis.

The second term in Eq.~(\ref{nrham}) is a non-Lorentz violating correction to the electron magnetic dipole moment. 
The most recent experimental measurement in Ref.~\cite{Odom:2006} finds
$g_e/2 = 1.00115965218085 (76)$. This is consistent with the theoretical calculation 
of Ref.~ \cite{Kinoshita:2006} which finds $g_e/2 = 1.00115965217586 \pm 8.48 \, 10^{-12}$. 
The second term in Eq.~(\ref{nrham}) gives a correction $\delta(g_e/2) =  \epsilon$, which
leads to the bound $\epsilon \leq 1.27 \, 10^{-11}$. This is slightly larger than the upper limit in Eq.~(\ref{epsrange}). 

The third term in Eq.~(\ref{nrham}) modifies the precession of the electron's spin in a background magnetic field.
The effect of this term is to shift the effective magnetic field seen by the electron spin:
\bea 
H &=& - \frac{g_e}{2} \mu_B \,\sigma \cdot \V{B}^\prime = - \frac{g_e}{2} \mu_B \, \sigma \cdot \V{B} 
 - \epsilon \, \mu_B \, (\V{n} \cdot \V{\sigma})\,  (\V{n} \cdot \V{B}) \nn \, ,
\eea 
where $\V{B}^\prime =  \,\V{B} + (2/g_e)\, \epsilon \, \V{n} \, (\V{n} \cdot \V{B})$,
so the the modified precession frequency is
\bea
 \omega = g_e \, \mu_B |\V{B}^\prime| &=& g_e\, \mu_B |\V{B}| \left(1 + \frac{2}{g_e}\,\epsilon (\V{\hat{n}} \cdot \V{\hat{B}})^2 \right) \, .
\eea
This will result in a time dependent precession frequency since $\V{B}$ is a magnetic field in a rotating 
frame on the surface of the Earth, while $\V{n}$ is a constant vector which is assumed to be fixed with 
respect to distant stars.  Assuming the $\V{B}$ field is oriented perpendicular  to the surface of the Earth and outward,
the magnitude of the daily variation in the precession frequency is
\bea\label{var}
\left|\frac{\Delta  \omega}{\omega_0}\right| = \frac{2}{g_e}\,\epsilon \left \{\begin{array}{ccr}
|\sin(2 \lambda) \sin(2\alpha) |& & \quad  |\tan \lambda  \cot \alpha | \geq 1 \\
{\it max}[\sin^2(\lambda \pm \alpha)] & &\quad | \tan \lambda  \cot \alpha | \leq 1
\end{array} \right. \, .
\eea
Here $\omega_0$ is the time-independent precession frequency in the absence of the third term in Eq.~(\ref{nrham}),
$\Delta  \,\omega$ is the maximal daily variation in $\omega$,
$\lambda$ is the latitude of the laboratory experiment, and $\alpha$ defines the angle between $\V{n}$ and
the axis of rotation of the Earth. The bound on $\Delta \,\omega$ from Ref.~{\cite{Mittleman:1999it} is 
$\Delta \,\omega \leq 1.2 $ Hz. The experiment is conducted with a $\V{B}$-field
of 5.85 T, corresponding to a precession frequency of $\omega_0 = 5.1 \, 10^{11}$ Hz. Assuming the geometrical factors in 
Eq.~(\ref{var}) are order unity we find $\epsilon \lesssim 10^{-11}$, which also exceeds the upper limit in Eq.~(\ref{epsrange}).

Next, we consider the effect of these terms on the atomic spectra of hydrogenic atoms applying perturbation theory in $\epsilon$. The factor
of $(\V{n} \times \V{E})\cdot \V{S}$ becomes $Z e \, \V{n} \cdot (\V{r} \times \V{S}) / r^3$ when the electric field of the atomic nucleus is
substituted in for $\V{E}$. In this form, we see that this term does not shift the energy spectrum to first order in $\epsilon$ as the matrix
element of this operator vanishes between any two states with the same $n$.
 We have
 \bea
[\V{p},H_0] = -i Ze^2 \frac{\V{r}}{r^3} \, ,\nn
\eea
where $H_0$ is the Coulomb Hamiltonian, so
\bea
Z e \, \frac{\V{n} \cdot ( \V{r} \times \V{S})}{r^3} = \frac{i}{e}[\V{n} \cdot ( \V{p} \times \V{S}),H_0] \, .
\eea
Therefore,
\bea
\langle n,l|Z e \, \frac{\V{n} \cdot ( \V{r} \times \V{S})}{r^3}|n,l'\rangle
= \langle n,l| \frac{i}{e}[\V{n} \cdot (\V{p} \times \V{S}),H_0]|n,l'\rangle = 0 \, .
\eea

The third new term in the non-relativistic Hamiltonian gives a rotationally noninvariant spin-orbit coupling because of the
magnetic field in the electron's rest frame generated by the rotating nuclear electric field. The correction to the
Hamiltonian  is 
\begin{equation*} 
\Delta H =  \epsilon \frac{Z \alpha}{2 m_e^2} \frac{1}{r^3} \V{L}\cdot \V{n} \, \V{S} \cdot \V{n}
\end{equation*}
The directional nature of this term causes a splitting in originally degenerate $2p_{3/2}$ states. Choosing 
the $\V{n}$ direction as the axis of quantization, and using
\begin{equation*}
\bra{n, l, m_l, m_s} \frac{1}{r^3} \ket{n, l, m_l, m_s} = \frac{Z^3 \alpha^3 m_e^3}{n^3 l (l + 1) (l + \frac{1}{2})}
\end{equation*}
we see that the originally degenerate energy levels split by
\begin{equation*}
E_{J_z = \pm \frac{3}{2}} - E_{J_z = \pm \frac{1}{2}} = \epsilon \frac{Z^4 \alpha^4 m_e}{72} = \epsilon \frac{Z^4 \alpha^2 (Ry)}{36}
\end{equation*}
This splitting must be less than the precision of the measurement of the $2s_{1/2} - 2p_{3/2}$ interval in hydrogen measured in 
Ref.~\cite{Hagley:1994}, otherwise two distinct lines would have been observed. Using the measured splitting frequency 
$9911.200 \pm .012$ MHz, we find $\epsilon < 2.5 \cdot 10^{-6}$.

Furthermore, this term causes a correction to the $n = 2$ Lamb shift: it leaves the $2s_{1/2}$ states unshifted while shifting the $2p_{1/2}$ states.
Using
\begin{align*}
\ket{2p, j = \frac{1}{2}, m_j = \frac{1}{2}} &= \sqrt{\frac{1}{3}} \ket{m_l = 0, +} - \sqrt{\frac{2}{3}} \ket{m_l = +1, -}\\
\ket{2p, j = \frac{1}{2}, m_j = - \frac{1}{2}} &= \sqrt{\frac{2}{3}} \ket{m_l = -1, +} - \sqrt{\frac{1}{3}} \ket{m_l = 0, -}
\end{align*}
We find  that the energy of $2p_{1/2}$ states shift  by
\begin{equation*}
\Delta E_{2p_{1/2}} = -\epsilon \frac{Z^4 \alpha^2 (Ry)}{72}
\end{equation*}
This is a correction to the measured Lamb shift, $|\Delta \omega_{Lamb}| = 1057.839 \pm .012$ MHz \cite{Hagley:1994}, which sets an upper bound for $\epsilon$ at $4.9 \, 10^{-6}$.

To summarize, bounds on Lorentz violating and Lorentz preserving couplings to magnetic fields from measurements of $g_e/2$ with trapped electrons yield
$\epsilon \leq 8.5 \, 10^{-12}$, which is just above the upper limit in Eq.~(\ref{epsrange}). Bounds from the spectroscopy of hydrogenic atoms are 
significantly weaker, with $\epsilon \lesssim 10^{-6}$. Bounds on a Lorentz violating electric dipole moment comparable to existing bounds on electron edm can rule out SIM(2) invariant masses.


\section{One loop self-energies \label{sec:loops}}

In this section we consider the one-loop corrections to the photon and electron propagators. Many high precision tests of
Lorentz invariance involve the propagation of light, for example, tests of cosmic birefringence. Therefore we would like to
know whether Lorentz violating modifications of the photon propagator arise in SIM(2) invariant theories. Since we have not 
added a SIM(2) invariant modification to the gauge field kinetic terms, such corrections must occur via loop effects in this theory. 
\footnote{It is possible to add a SIM(2) invariant
modification of the gauge field kinetic term~\cite{Lindstrom:2006xh}:
\bea
 \frac{m_\gamma^2}{2} \frac{1}{i n \cdot D} n_\alpha F^{\alpha \mu} \frac{1}{i n \cdot D} n_\beta F^{\beta}_\mu
\, .
\eea
In $n\cdot A = 0 $ gauge this is simply a photon mass term. The present bound on the photon mass is
$m_\gamma < 6 \, 10^{-17}$ eV~\cite{PDG:2006}, which is about 15 orders of magnitude smaller than mass scale set by neutrino mass 
differences. For this reason we have chosen to ignore such a term.} 
It is also important to understand how the nonlocality of the SIM(2) invariant action 
affects the calculation of loops. Straightforward
evaluation of one loop corrections to the photon and electron propagators results in unregulated infrared
(IR) divergences. These are due to the factor of $1/(n \cdot p)$ in the electron propagator. These IR divergences cannot 
be regulated by dimensional regularization or by taking external legs in the diagrams off-shell, as is the case  
for IR divergences that arise in local quantum field theory diagrams. This is not so surprising given the
nonlocality of the fermion action. It indicates that the theory is ill-defined at the quantum level, unless a 
prescription for dealing with these IR singularities is given. 

Since the physical origin of the nonlocality is unknown, it is not clear how to construct an adequate
prescription. Motivated by the observation that the action is ill-defined for off-shell modes with 
$n\cdot p =0$, we suggest that these modes be excluded from the theory altogether. Since on-shell momenta
cannot satisfy $n\cdot p = 0$, these modes only appear in internal lines in Feynman diagrams. We propose that  in loop diagrams
$n\cdot p = 0$ modes should not be allowed to propagate. This can be implemented by imposing  a ``+'' prescription for
defining the quantum mechanical theory, in which loop integrals over the $n\cdot l$
component of loop momenta $l$  are {\it defined} to  be
\bea
\int d n\cdot l \, \frac{f(n\cdot l)}{n \cdot l} 
&\longrightarrow &  \int d n\cdot l\, \frac{f(n\cdot l) }{[n \cdot l]_+} \nn \\
&=& \int d n\cdot l \, \frac{f(n\cdot l)-f(0)}{n \cdot l} 
\eea
This prescription tames the IR singularities. The novel contribution to the photon self  energy in the
SIM(2) invariant theory vanishes on-shell with this prescription. At one-loop, the photon propagator is essentially
unmodified relative to the usual self-energy correction in QED. There are no Lorentz violating effects and
no photon mass is induced. This last fact is important since it is possible to write down a SIM(2)
invariant, gauge invariant photon mass term (see footnote). If an $O(\alpha \, m_\beta)$ mass term were
induced by radiative corrections the theory would be phenomenologically unviable. The ``+'' prescription 
also removes IR divergences in the electron self-energy.  Moreover, in that case we find that the radiative 
correction to the SIM(2) invariant mass term is {\it finite}. This is something that one might expect from the naive degree of
divergence of the SIM(2) invariant correction to the electron propagator. Counting the naive degree  of
divergence in higher order diagrams suggests that the finiteness of the correction to the SIM(2)
invariant  mass term will persist in higher orders in perturbation theory. If this is the case, then
SIM(2) masses are  {\it natural} in the technical sense. To have small SIM(2) invariant mass terms, 
presumably of order the neutrino mass,  will not require fine tuning at each order in perturbation theory.
However, the question of why the scale of SIM(2)  invariant masses is so much smaller than the other mass
scales in nature remains unanswered.

We begin by computing the electron self-energy in $n\cdot A=0$ gauge. The diagram is the same as in 
QED and the result for one-loop self-energy is 
\bea\label{ese}
-i \Sigma(p) &=& e^2 \int \frac{d^D q}{(2\pi)^D}\frac{\gamma_\mu 
\left( - \qsl + m_0 +\frac{m_\beta^2}{2} \frac{\ns}{n \cdot q} \right)\gamma_\nu}{(q+p)^2 \,(q^2-m_e^2)}
\left(-g^{\mu \nu} + \frac{n^\mu \, (p+q)^\nu  + (p+q)^\mu \, n^\nu}{n\cdot (p+q)}\right) \nn \\
&=& e^2 \int \frac{d^D q}{(2\pi)^D}\frac{\gamma_\mu 
\left( - \qsl + m_e  \right)\gamma_\nu}{(q+p)^2 \,(q^2-m_e^2)}
\left(-g^{\mu \nu} + \frac{n^\mu \, (p + q)^\nu  + (p + q)^\mu \, n^\nu}{n\cdot (p + q)}\right) \nn \\
&+& e^2 (D-2) \frac{m_\beta^2}{2 m_e}\int \frac{d^D q}{(2\pi)^D}\frac{1}{(q+p)^2 \,(q^2-m_e^2)}\nn \\
&+& e^2 (D-2) \frac{m_\beta^2}{2 }\int \frac{d^D q}{(2\pi)^D}\frac{\ns}{(q+p)^2 \,(q^2-m_e^2)\, n\cdot q} 
\eea
We have routed momenta so that $-q$ is the momentum going through the electron propagator.
In the second line of Eq.~(\ref{ese}), we have expanded $m_0 = m_e -m_\beta^2/(2 m_e)$. There are three terms
in the electron self-energy. The first is the usual electron self-energy in $n\cdot A=0$ gauge. The 
second term is Lorentz invariant and gives a shift in the electron mass of order $\alpha \, m_\beta^2/m_e \log(m_e/\mu)$.
Since the integral in the second term is divergent, it gives
a tiny $O(m_\beta^2/m_e^2)$ correction to the anomalous dimension of the electron mass.
The third term violates Lorentz invariance, preserves SIM(2) invariance and contains IR divergences
due to the nonlocality of the theory. We will focus on the evaluation of the last integral in Eq.~(\ref{ese}).

Using light-cone coordinates where $n\cdot q = q^-$ and $q^2=q^+q^- - \V{q}^2$,
and restoring the $i\epsilon$'s, we find
\begin{align*}
I &\equiv \int \frac{d^Dq}{(2 \pi)^D} \, \frac{1}{n \cdot q (q^2 - m_e^2 + i \epsilon) ((p+q)^2 + i \epsilon)}\\
&= \int \frac{d^{D-2} \V{q} dq^+ dq^-}{2(2 \pi)^D} \frac{1}{ q^- (q^+ q^- - \V{q}^2 - m_e^2 + i \epsilon) ((p^- + q^-) (p^+ + q^+)
-(\V{p}+\V{q})^2+i\epsilon)}
\end{align*}
where the boldface vectors indicate spatial coordinates orthogonal to both light-cone coordinates. 
We evaluate the $q^+$  integral using contour integration. The poles in the complex $q^+$ plane have to 
lie on opposite sides of the real axis in order for the contour integral in the $q^+$ plane not to 
vanish. Therefore, if we close the contour around  the pole at $q^+ =(\V{q}^2+m_e^2)/q^-$, 
the remaining $q^-$ integral is constrained to lie within the range $- p^- < q^- <0$, if $p^- >0$,
or  $0 < q^- < |p^-|$, if $p^- <0$. Without loss of generality, we assume $p^- >0$ and close the contour 
around the pole at $q^+ =(\V{q}^2+m_e^2)/q^-$. Then we shift $\V{q}$ to complete the square in the 
denominator and define the variable, $x$,
\bea\label{xdef}
x = \frac{-q^-}{p^-} \, .
\eea
In terms of this variable, the result can be written as 
\bea\label{esans}
I &=& - \frac{i}{n \cdot p} \int_0^1 \frac{dx}{x} \int \frac{d^{D-2} \V{q}}{2(2 \pi)^{D-1}} \frac{1}{\V{q}^2 + (1 - x) m_e^2 - x(1 - x) p^2} \nn \\
&=& - \frac{i}{16\pi^2 n \cdot p} \int_0^1 \frac{dx}{x} \left(\frac{1}{\hat  \epsilon} - \log\left(\frac{(1-x) m_e^2-x(1-x) p^2}{\mu^2}\right)\right)
\eea
where we evaluated the integral over $\V{q}$ in dimensional regularization with $D=4 -2 \epsilon$ and $1/\hat{\epsilon} \equiv 
1/\epsilon -\gamma_E + \log(4 \pi)$. The $x$ integral is infrared divergent. This divergence is not regulated by dimensional
regularization and cannot be removed  by taking the electron off-shell. From our definition of $x$, it is clear that the origin of
this IR divergence is from the limit $n\cdot q \to 0$. In this limit the Lorentz violating  term in the electron propagator blows
up. Rather than using contour integration to evaluate the integral we could have combined the denominators using Feynman parameters, 
then we would have obtained an expression identical to Eq.~(\ref{esans}), where $x$ in this evaluation corresponds to  Feynman
parameter. In this way of doing the integral the  physical origin of the IR divergence is somewhat obscure.

The ``+''-prescription for defining the SIM(2) invariant theory deals with the IR divergence by replacing the pole in the 
electron propagator with the +-distribution:
\bea\label{pp}
\frac{1}{n\cdot q} \to \left(\frac{1}{n \cdot q}\right)_+ \, .
\eea
The integral becomes 
\bea
I &=& - \frac{i}{16\pi^2 n \cdot p} \int_0^1 dx \left[\frac{1}{x} \right]_+\left(\frac{1}{\hat  \epsilon}
 - \log\left(\frac{(1-x) m_e^2-x(1-x) p^2}{\mu^2}\right)\right) \nn \\
&=& \frac{i}{16 \pi^2 n \cdot p} \int_0^1 \frac{dx}{x} \log\left(1 - x - x(1 - x) \frac{p^2}{m_e^2}\right) \, ,
\eea
so the correction to $-i \Sigma(p)$ coming from the third term is 
\bea
\Delta\Sigma(p) = \frac{\alpha}{4 \pi} m_\beta^2 \cdot \frac{\ns}{n \cdot p} 
\left(\frac{\pi^2}{6} + {\rm Li}_2\left(\frac{p^2}{m_e^2}\right)\right)\,.
\eea
Note that now the correction is both UV and IR finite. Going on the mass shell, $p^2 = m_e^2$, we find the following 
correction to the SIM(2) invariant electron mass;
\bea
\delta m_\beta^2 = \frac{\alpha \pi}{6} m_\beta^2 \,.
\eea
As stated earlier, this correction is finite.

Next we evaluate the one-loop corrections to the photon propagator. 
The one-loop diagrams are shown in Fig.~\ref{fig:phoprop}, the Feynman rules needed to evaluate these diagrams are
derived in the Appendix. Both diagrams
are necessary to obtain a result that satisfies the QED Ward identity, $p_\mu \Pi^{\mu \nu}(p) =0$.
\begin{figure}[!h]
   \centering
   \includegraphics[width=0.7\textwidth]{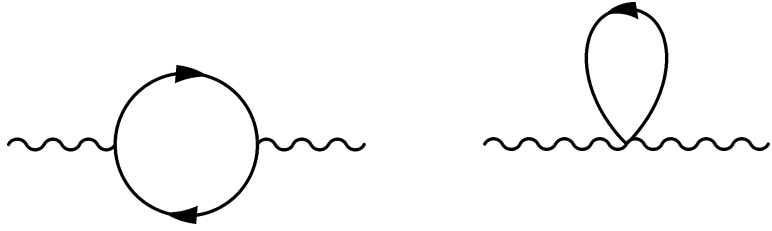}
   \caption{One-loop diagrams for the photon self-energy in VSRQED.}
   \label{fig:phoprop}
\end{figure}
There are $O(m_\beta^2)$ corrections to both the vertices and the electron propagators.
Keeping all terms $O(m_\beta^2)$ terms, and denoting the $O(m_\beta^2)$ correction to the 
photon self-energy as $\Delta \Pi^{\mu \nu}$ we find
\bea\label{phopol}
i \Delta \Pi^{\mu \nu} &=& -2 e^2 m_\beta^2\left(-g^{\mu \nu} + \frac{n^\mu p^\nu + p^\mu n^\nu}{n\cdot p}
- \frac{p^2}{(n\cdot p)^2} n^\mu n^\nu\right) \times \nn \\
&& \int \frac{d^D l}{(2 \pi)^D} \frac{1}{l^2-m_e^2}\frac{1}{(l+p)^2-m_e^2}
\left(\frac{n\cdot p}{n \cdot l} - \frac{n\cdot p}{n \cdot (l+p)}\right) \, .
\eea
Changing variables $l \to -l-p$ shows that the two integrals in Eq.~(\ref{phopol}) are identical.
The integrals appearing in the photon polarization are essentially identical to those  appearing
in the electron self-energy and are IR divergent due to the poles in $n\cdot l$ and $n \cdot (l+p)$.
Note that in this case these factors come from both the electron propagator and the SIM(2) invariant
modifications to the electron photon coupling. Applying the ``+'' prescription yields:
\bea
\Delta \Pi^{\mu \nu} = - \frac{\alpha}{\pi} m_\beta^2
\left(-g^{\mu \nu} + \frac{n^\mu p^\nu + p^\mu n^\nu}{n\cdot p}
- \frac{p^2}{(n\cdot p)^2} n^\mu n^\nu\right)  \int_0^1 \frac{dx}{x} \log\left(1 - x(1 - x) \frac{p^2}{m_e^2}\right)
\eea
This integral can be evaluated in terms of dilogarithms. It clearly vanishes on-shell when $p^2=0$. 
There is no photon mass term generated in C invariant VSRQED at one loop, and no Lorentz violating
modification of light propagation. 

Divergences of the type encountered in the SIM(2) invariant theory also arise in ordinary gauge theory
when quantized in $n\cdot A = 0$ gauge. In this case the gauge field propagator has poles in $1/(n\cdot p)$
and a prescription is required to define loop integrals~\cite{Leibbrandt:1987}. The prescription 
proposed in Ref.~\cite{Leibbrandt:1987} is to define loop integrals by making the following substitution:
\begin{equation*}
\frac{1}{n \cdot q} \rightarrow \frac{\bar{n} \cdot q}{(n \cdot q)(\bar{n} \cdot q) + i \epsilon}
\end{equation*}
where $\bar n^\mu$ is defined to be $(1,-\V{n})$ if $n^\mu = (1,\V{n})$.
This prescription yields sensible results for gauge theories quantized in light-cone gauge but 
will not work in the present case because it violates SIM(2) invariance.  For example, this prescription 
leads to the  following integral:
\begin{equation*}
\int \frac{d^D q}{(2 \pi)^D}\, \frac{\ns}{(q - p)^2 q \cdot n} = \ns \frac{i}{(4\pi)^2}
\frac{2 p \cdot \bar{n}}{n \cdot \bar{n}} \frac{1}{\hat{\epsilon}} + ... \, .
\end{equation*}
The left-hand side is SIM(2) invariant, but the right-hand side is not, because the four-vector $\bar{n}^\mu$ is not left
invariant by all of the generators in the little group of $n^\mu$.

The prescription developed here for handling IR divergences is clearly ad-hoc but was forced upon us by the 
necessity of obtaining sensible results from one-loop calculations in the SIM(2) invariant theory. Our proposal
is only a preliminary attempt to try to address this problem. It is satisfactory that  one-loop  corrections
to the photon and electron propagators are well-defined, free of IR divergences and for the photon propagator
respect the QED Ward identity. Interestingly, the corrections to the electron mass is finite, suggesting 
that if this prescription is valid a SIM(2) invariant mass term is technically natural. Several questions remain unanswered. 
Will the ``+''-prescription continue to yield sensible results in higher order diagrams? Will the 
``+''-prescription respect gauge invariance? Though the QED Ward identity for the photon propagator is satisfied at one 
loop, we have not shown that Ward identities will be respected in higher order diagrams or for other Green's functions. 
Are there other sensible prescriptions and do they give different physical results? Clearly, these issues
must be addressed before we can be satisfied that the SIM(2) invariant theory is well-defined at the quantum level.

\section{Lepton Family Number Violation: $\mu \rightarrow e + \gamma$}

In this section we consider the decay 
$\mu \to e + \gamma$, which is allowed because the SIM(2) invariant  mass terms in the lepton 
sector are not family diagonal. We will not impose C invariance but consider the more 
general Lagrangian in Eq.~(\ref{genL}).
There is a direct $\mu \to e + \gamma$ coupling  from the covariant 
derivative in the off-diagonal SIM(2) mass term. This coupling vanishes 
in $n\cdot A$ = 0 gauge. There are also two diagrams which arise due to 
$\mu \to e$ transitions induced by the off-diagonal SIM(2) invariant mass term combined 
with an ordinary QED photon coupling to the electron or muon.
The Feynman rule for the $\mu \to e$ transition and the decay diagrams are shown in Fig.~\ref{fig:muegamma}. 
The factors $M^L_{ij}$ and $M^R_{ij}$ in Fig.~\ref{fig:muegamma} are given by
\bea
M^L_{ij} = (V_M m_\nu^2 V_M^\dagger)_{ij} \qquad M^R_{ij} = (U_R^\dagger \tilde M^2 U_R)_{ij} \, .\nn
\eea
The left-handed couplings are quite small because they are proportional to differences of neutrino masses squared 
since the off-diagonal mass terms vanish in the limit of degenerate neutrinos. 
For example, using the parametrization of the neutrino mixing matrix in the PDG~\cite{PDG:2006},
for $M^L_{12}$ we find 
\bea\label{bemu}
M^L_{12} = (V_M m_\nu^2 V_M^\dagger)_{12} &=& s_{12} \, c_{12}\, c_{13}\, c_{23}\, \Delta m_{21}^2 
+ e^{- i \delta} s_{13} \, s_{23}\, c_{1 3} (\Delta m_{31}^2 -s_{12}^2\, \Delta m_{21}^2) \nn \\
&=&\frac{1}{\sqrt{2}} \, s_{12} \, c_{12}\, \Delta m_{21}^2+ 
\frac{1}{\sqrt{2}} \, e^{-i \delta} s_{13}\,\left(\Delta m_{31}^2-s_{12}^2\,\Delta m_{21}^2\right)
\, .
\eea
In the second line we make the approximation $s_{23} \approx c_{23} \approx 1/\sqrt{2}, c_{13} \approx 1$.  From 
the current values in the PDG~\cite{PDG:2006}, we know that the magnitude of the 
first term in the last line of the Eq.~(\ref{bemu}) is $2.6 \, 10^{-5} \,{\rm eV}^2$ while the
magnitude of the second term is in the range $3.0-4.7 \, 10^{-4} \,{\rm eV}^2$. The relative phase
of the two terms is unknown.
Assuming that the decay is saturated by the contribution from the left-handed SIM(2) mass terms we obtain lower bounds for 
${\rm Br}[\mu \to e + \gamma]$ that are $12$ to $15$ orders of magnitude below the current experimental
bounds. On the other hand, the right handed couplings are completely undetermined and so the 
the current bound on ${\rm Br} [\mu \to e +\gamma]$ gives a nontrivial constraint on 
$(U_M^\dagger \tilde M^2 U_M)_{12}$.

\begin{figure}[!t]
   \centering
   \includegraphics[width=1.0\textwidth]{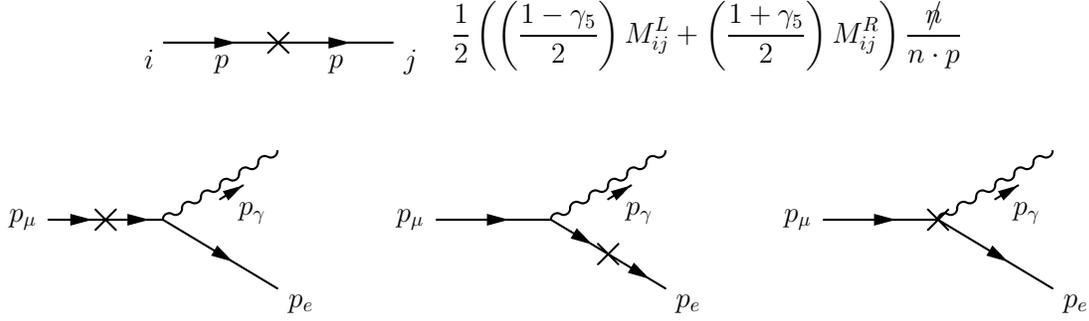}
   \caption{Feynman rule for fermion flavor changing vertex and associated amplitudes for $\mu^- \rightarrow e^- \gamma$ at tree level. 
   The third diagram comes from the expansion $\frac{1}{n \cdot D}$ term in powers of $A$ and does not appear in $n \cdot A = 0$ gauge.}
   \label{fig:muegamma}
\end{figure}

For the differential decay rate, we find 
\bea
\frac{d \Gamma[\mu \to e +\gamma]}{d \Omega} = \frac{\alpha}{16 \pi m_\mu^3}
\left(\frac{(1+x^2)(|M^L_{12}|^2+ |M^R_{12}|^2) -4 \, x \, {\rm Re}(M^L_{12} \, M^{R*}_{12})}{(1-x^2)}\right)
\frac{(n\cdot q)^2}{n\cdot p_e \,n\cdot p_\mu}
\eea
where $q$, $p_e$, and $p_\mu$ are the momentum of the photon, electron, and  muon, respectively.
We have used the ordinary QED wavefunctions and propagators for the electron and 
muon because the  SIM(2) invariant corrections to these give contributions which are suppressed by more powers
of $m_\nu^2$ that can be neglected.
Lorentz violation manifests itself in the angular dependence of the differential decay rate, which 
vanishes when the photon is emitted along the direction defined by $n^\mu$.
Note the result is invariant under rescaling of $n^\mu$ as required by SIM(2) invariance. 
This allows to one to choose $n^\mu = (1,\V{n})$ where $\V{n}$ is a unit vector. Performing the angular
integral is straightforward and the result for the total decay rate is
\bea
\Gamma[\mu \to && \!\!\!\! e + \gamma] = \\
&&\frac{\alpha}{4 m_\mu^3}
\frac{(1+x^2)(|M^L_{12}|^2+ |M^R_{12}|^2) -4 \, x \, {\rm Re}(M^L_{12} \, M^{R*}_{12})}{(1-x^2)^2}
 \left( \log\left(\frac{1}{x^2}\right) - \frac{3}{2}+2 x^2 - \frac{x^4}{2}\right)  , \nn
\eea
where $x = m_e/m_\mu$. Since $x \approx 0.005$, if $|M^R_{12}| < 50 \, |M^L_{12}|$, then the combination of $|M^R_{12}|^2 
- 4 x {\rm Re}(M^L_{12} \, M^{R*}_{12})$ will be positive definite. In this case, the contribution from left-handed SIM(2) mass terms
can be used to establish a lower bound on the branching ratio for $\mu \to e + \gamma$, which we find to be
\bea
{\rm Br}[\mu \to e + \gamma] \geq   3.2 \, 10^{-26} - 1.0 \, 10^{-23} \, .
\eea
The first number corresponds to $s_{13}=0$ while the second corresponds to $s_{13}$ 
equal to its current upper bound. These branching ratios are a factor of 
$10^{-15}$-$10^{-12}$ smaller than the current experimental 
bound, ${\rm Br}[\mu \to e + \gamma] < 1.2 \, 10^{-11}$~\cite{PDG:2006}, and therefore uninteresting.

The smallness of this branching ratio when $M^L_{12}$ dominates the decay
is directly related to the smallness of the 
neutrino mass squared differences, $\Delta m_{ij}^2$. $M^R_{12}$ will be proportional to  the mass squared 
differences
of the eigenvalues of $\tilde M^2$ which are the SIM(2) invariant contributions to the right handed lepton masses.
Earlier we saw that $g_e - 2$ experiments with trapped electrons constrained the flavor diagonal components:
$(V_M m_\nu^2 V_M^\dagger)_{11} \approx (U_R^\dagger \tilde M^2 U_R)_{11}$. The nonobservation of $\mu \to e + \gamma$
puts a constraint on the off-diagonal component,  $M^R_{12}$. 
We find $M^R_{12} = (U_R^\dagger \tilde M^2 U_R)_{12} \leq 0.50 \, 10^3\, {\rm eV}^2$. Note that
$0.5 \, 10^3 \, {\rm eV}^2 = (22.4 \, {\rm eV})^2$, so the upper limit on $M^R_{12}$
implied by $\mu \to e + \gamma$ is significantly larger than the bound obtained for the flavor diagonal term,
$M^R_{11}$.

\section{Summary}

In this paper we have studied the implications of $SU(2)_L \times U(1)$ gauge symmetry for the SIM(2) invariant neutrino mass 
proposed in Ref.~\cite{Cohen:2006ir}. This mass term induces Lorentz violating effects in QED as well as
tree-level lepton family number violating  interactions. Tests of Lorentz invariance in experiments on trapped electrons
severely constrain a possible C violating SIM(2) invariant mass term in QED. We studied VSRQED, which contains a C invariant
SIM(2) invariant mass term for the electron. We derived a  Hamiltonian for nonrelativistic electrons interacting with
background electric and magnetic fields and used it to  determine the impact of VSRQED on $g_e-2$ measurements with trapped
electrons as well as the spectroscopy of hydrogenic atoms. Corrections from this Hamiltonian are suppressed by $\epsilon =
m_\beta^2/m_e^2$. This parameter is expected to lie within the range $9.2 \, 10^{-17} \leq \epsilon \leq 3.8 \, 10^{-12}$
based on our present understanding of neutrino masses. Experiments on trapped electrons place limits on $\epsilon$ that lie
just above the upper limit of this range, while constraints from the spectroscopy of Hydrogenic atoms give weaker bounds on
$\epsilon$. Bounds on a Lorentz violating electric dipole moment comparable in size to existing bounds on the electron
electric dipole moment could falsify the  SIM(2) invariant neutrino mass hypothesis. We found the tree level rate for
$\mu \to e + \gamma$ is many orders of magnitude below current experimental limits.

We also examined one loop self-energy diagrams in VSRQED. These suffer from unregulated IR divergences due to the nonlocality
of the SIM(2) invariant action, indicating that the theory is ill-defined at the quantum level. We suggested a prescription
for defining loop integrals which gives sensible results for the one-loop diagrams considered in this paper. QED Ward
identities are satisfied, no Lorentz violating effects in the photon propagator are induced, and the one-loop correction to
the electron's SIM(2) invariant Lorentz violating mass is finite.  Future work on SIM(2) invariant  theories needs to address
whether this prescription, or some other method of defining loop integrals, will give satisfactory results to all orders and
for all Green's functions.

\acknowledgments

T.M.is supported in part by the U.S. Department of
Energy under Grants No.~DE-FG02-05ER41368-0. DE-FG02-05ER41376, and DE-AC05-84ER40150.
A.D. is supported by the National Science Foundation under grants 
DMS-0074072 and DMS-0301476. We thank R. Plesser for useful discussions.

\appendix

\section{Feynman Rules for  VSRQED}

The Lagrangian for VSRQED is 
\bea
{\cal L} = \bar\psi \left(i \Dsl - m_0 - \frac{1}{2} m_\beta^2 \frac{\ns}{i n\cdot D} \right) \psi - \frac{1}{4} F^{\mu \nu}F_{\mu \nu} \, ,
\eea
where $D_\mu = \partial_\mu - i e A_\mu$ is a covariant derivative. There is no modification to the gauge field kinetic terms. The parameter
$m_\beta^2$ is $(V_M m_\nu^2 V_M^\dagger)_{11} = \sum_i |V_{ei}|^2 m_{\nu,i}^2$. This is the so-called electron-based neutrino mass squared.  The nonlocal SIM(2) invariant
mass term gives the correction to the propagator in Eq.~(\ref{cinv}) and also gives rise to additional couplings of electrons and photons. 
The additional couplings are absent if the theory is quantized in light-cone axial ($n\cdot A=0$) gauge. In many cases, it is simplest to use
light-cone   gauge when analyzing this theory. Of course, to check that gauge invariance is satisfied we need to be able to calculate in 
arbitrary gauges. We will provide Feynman rules in both light-cone  gauge and covariant gauges in this Appendix. We also check our covariant gauge
Feynman rules by verifying that the tree level VSRQED Compton scattering amplitude satisfies the Ward identity.

The propagators for the electron and photon in light-cone gauge are shown in Fig.~\ref{fig:frules},
and the electron-photon coupling is standard. The Feynman rules are simplest in this gauge.
\begin{figure}[!t]
   \centering
   \includegraphics[width=1.0\textwidth]{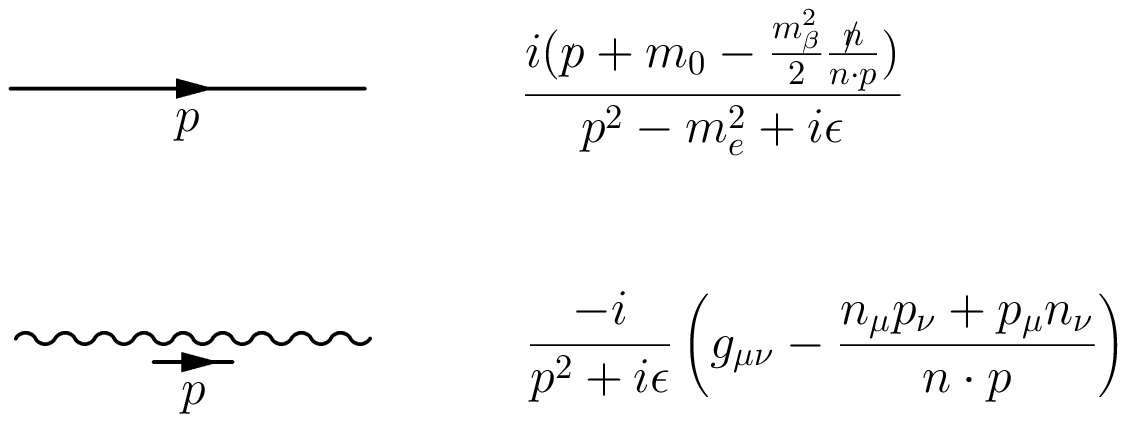}
   \caption{Light cone gauge Feynman rules in VSRQED. Unlisted rules are unmodified from standard QED.}
   \label{fig:frules}
\end{figure}

In covariant gauges, the photon propagator is standard and the electron propagator remains the same as in light-cone gauge, 
but there are now an infinite number of interactions due to the $\frac{1}{i n \cdot D}$ term. We can make use of Wilson lines to express these terms in a form that is convenient for deriving the Feynman rules. 
The Wilson line, 
\begin{equation*}
W(x) = \exp\left(- i e \int_{-\infty}^0 d\lambda\, n \cdot A(x + n \lambda)\right) \, ,
\end{equation*}
satisfies the equation $i n \cdot D W(x) = 0$ and is a unitary operator. As an operator equation we have
\begin{align*}
W^\dagger(x) n \cdot D W(x) &= W^\dagger(x) \left[n \cdot D W(x)\right] + W^\dagger(x) W(x) n \cdot \d\\
&= n \cdot \d \, ,
\end{align*}
which leads to \cite{Bauer:2001yt}
\begin{equation}
\frac{1}{i n \cdot D} = W(x) \frac{1}{i n \cdot \d} W^\dagger(x) \label{eq:WLD} \, .
\end{equation}
Expanding the right-hand side in powers of $A$ will give the Feynman rules. In order to expand $W(x)$ we need the intermediate integral
\begin{align*}
-i e \int_{-\infty}^0 d\lambda\, n \cdot A(x + n \lambda) &= -i e \int \dfs^4p\, n \cdot \tilde{A}(p) \int_{-\infty}^0 d\lambda\, e^{-i p \cdot (x + n \lambda)}\\
&= -i e \int d^4p\, n \cdot \tilde{A}(p) e^{-i p \cdot x} \cdot \frac{i}{n \cdot p + i \epsilon} \, .
\end{align*}
So expanding $\eqref{eq:WLD}$ to first order in $\tilde{A}$ in momentum space (remembering that $i \d$ in momentum space will become 
the sum of the momenta of all operators to its right), we find
\begin{align*}
W \frac{1}{i n \cdot \d} W^\dagger \rightarrow \frac{e \, n \cdot \tilde{A}(q)}{n \cdot q} \frac{1}{n \cdot p} - \frac{1}{n \cdot (p + q)} \frac{e\,  n \cdot \tilde{A}(q)}{n \cdot q}
&= \frac{e \, n \cdot \tilde{A}(q)}{n \cdot q} \left(\frac{1}{n \cdot p} - \frac{1}{n \cdot p'}\right)\\
&= \frac{e \, n \cdot \tilde{A}(q)}{n \cdot q} \frac{n \cdot(p' - p)}{(n \cdot p)(n \cdot p')}\\
&= \frac{e \, n \cdot \tilde{A}(q)}{(n \cdot p)(n \cdot p')} \, .
\end{align*}
Collecting this term with the traditional QED interaction term, the QED vertex is modified as shown in 
Fig.~\ref{fig:frules2}. The structure of this vertex is very similar to the  
modification of the weak current derived in the analysis of Ref.~\cite{Cohen:2006ir}. In addition to the modification
of the QED vertex, there are also additional vertices with arbitrary numbers of photon fields.
For example, second order term in $\tilde{A}$ gives
\begin{align*}
e^2 \, n \cdot \tilde{A}(q_1) & n \cdot \tilde{A}(q_2) \Big(\frac{1}{n \cdot q_1} \frac{1}{n \cdot q_2} \frac{1}{n \cdot p} - \Big(\frac{1}{n \cdot q_1} \frac{1}{n \cdot (p + q_2)} \frac{1}{n \cdot q_2} +\\
&\frac{1}{n \cdot q_2} \frac{1}{n \cdot (p + q_1)} \frac{1}{n \cdot q_1}\Big) + \frac{1}{n \cdot (p + q_1 + q_2)} \frac{1}{n \cdot q_1} \frac{1}{n \cdot q_2}\Big)\\
&= \frac{e^2 \, n \cdot \tilde{A}(q_1) n \cdot \tilde{A}(q_2)}{(n \cdot q_1)(n \cdot q_2)} \left(\frac{1}{n \cdot p} - \left(\frac{1}{n \cdot (p + q_1)} 
+ \frac{1}{n \cdot (p + q_2)}\right) + \frac{1}{n \cdot p'}\right) \, ,
\end{align*}
which leads to the two-photon vertex shown in Fig.~\ref{fig:frules2}.

\begin{figure}[!t]
   \centering
   \includegraphics[width=1.0\textwidth]{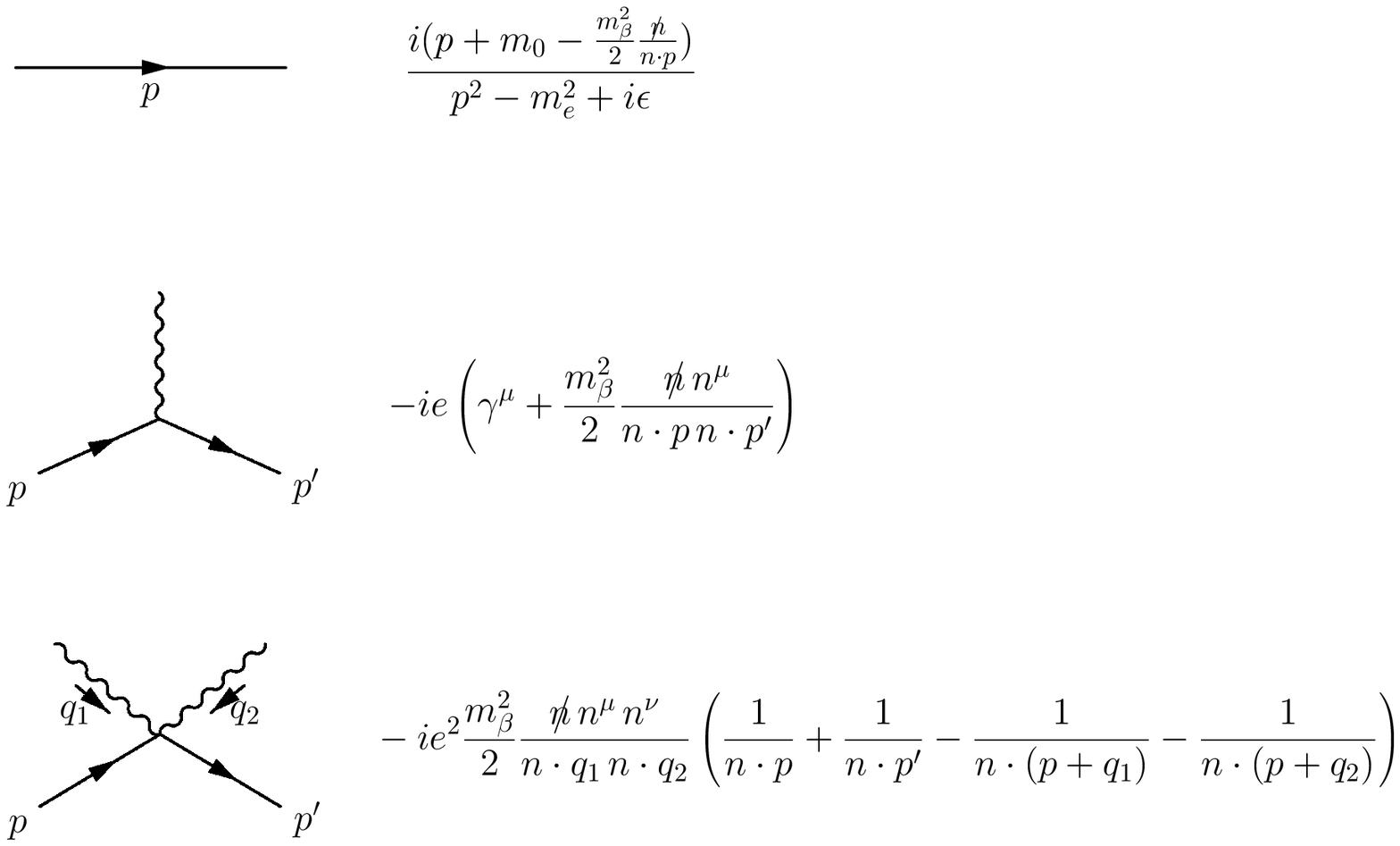}
   \caption{Covariant gauge Feynman rules in VSR QED up to two powers of $A$. Unlisted rules are unmodified from standard QED.}
   \label{fig:frules2}
\end{figure}

We can check these Feynman rules by verifying that the tree level amplitude for Compton scattering obeys the Ward identity. 
The VSRQED amplitude for Compton scattering at tree level in Feynman gauge is a sum of three diagrams
shown in Fig.~\ref{fig:csfeyn}. We define the incoming electron and photon momentum to be $p$ and $q$, respectively, 
and define the  outgoing  electron and photon momentum to be $p'$ and $q'$, respectively. Next define

\begin{figure}[tb]
  \centering 
  \includegraphics[width=1.0\textwidth]{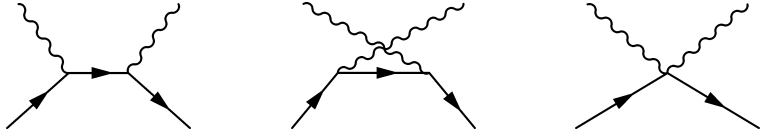}%
  \caption{Leading order Compton scattering diagrams in VSRQED. 
  }
  \label{fig:csfeyn}
\end{figure}

\bea
n_p^\mu \equiv p^\mu - \frac{m_\beta^2}{2} \frac{n^\mu}{n\cdot p} \, 
\eea
and 
\bea
\Gamma_\mu(k,k') \equiv \gamma_\mu + \frac{m_\beta^2}{2}\frac{\ns \, n_\mu}{n\cdot k \, n\cdot k'} \, .
\eea
Note that
\bea\label{t1} 
n_p^2 - m_0^2 = p^2 - m_\beta^2 -m_0^2 &=& p^2 -m_e^2 \nn  \\
(\ns_p - m_0) u_n(p) &=& 0 \, . 
\eea 
The first two diagrams in Fig.~\ref{fig:csfeyn} then contribute
\bea\label{ft}
i \mathcal{M} &=& -i e^2 \bar u_n(p')\left(
\Gamma_\nu(p+q,p')\, \frac{\ns_{p+q}+m_0}{n_{p+q}^2 -m_0^2}\, \Gamma_\mu(p+q,p)  \right. \nn \\
&&\left. + \Gamma_\mu(p'-q,p')\, \frac{\ns_{p'-q}+m_0}{n_{p'-q}^2 -m_0^2}\, \Gamma_\nu(p-q',p)\right) 
u_n(p)\,  \epsilon^{*\nu}(q') \epsilon^\mu(q) \,  .
\eea
Here $\epsilon^\mu$ and ${\epsilon^*}^\nu$ are
photon polarization vectors. The Ward identity demands that the amplitude vanish under
the replacement $\epsilon^{\mu}(q) \to q^{\mu}$. Making this substitution in Eq.~(\ref{ft})
and using the identities
\bea
q^{\mu} \Gamma_\mu(p+q,p) &=& \ns_{p+q} - \ns_p \nn \\
q^{\mu} \Gamma_\mu(p'-q,p') &=& \ns_{p'} - \ns_{p'-q}\, ,
\eea 
along with the identities in Eq.~(\ref{t1}), it is easy to show that the result is
\bea\label{wift}
i \mathcal{M}[\epsilon^{\nu}(q) \to q^{\nu}] &=& -i e^2 \bar u_n(p')\bigg(
\Gamma_\nu(p+q,p')-\Gamma_\nu(p-q',p)\bigg) u_n(p)\, \epsilon^{*\nu}(q')  \\
&=& -i e^2 \frac{m_\beta^2}{2} \bar u_n(p') \, \ns \, n_\nu \,
\left( \frac{1}{n\cdot (p+q) \, n\cdot p'} -  \frac{1}{n\cdot (p-q') \, n\cdot p}\right) 
u_n(p) \, \epsilon^{*\nu}(q')\, . \nn
\eea
The third diagram in Fig.~\ref{fig:csfeyn} gives the following contribution:
\bea
i \mathcal{M}[\epsilon^{\nu}(q) \to q^{\nu}] &=& -i e^2 \frac{m_\beta^2}{2} 
\bar u_n(p') \frac{\ns \, n_\nu}{n \cdot q'} 
\left(\frac{1}{n \cdot p} + \frac{1}{n \cdot p'} - \frac{1}{n \cdot (p'+q')} -\frac{1}{n \cdot p-q'} \right)
 u_n(p) \, \epsilon^{*\nu}(q')\nn \\
 &=& i e^2 \frac{m_\beta^2}{2} 
\bar u_n(p') \, \ns \, n_\nu  
\left(\frac{1}{n\cdot (p+q) \, n\cdot p'} -  \frac{1}{n\cdot (p-q') \, n\cdot p} \right)
u_n(p)\, \epsilon^{*\nu}(q') \, ,
\eea
which exactly cancels Eq.~(\ref{wift}), so the Ward identity is satisfied.

\end{document}